\overfullrule=0pt
\input harvmac
\def\a{{\alpha}}

\def\ah{{\widehat \a}}
\def\lh{{\widehat \lambda}}

\def\wh{{\widehat w}}

\def\l{{\lambda}}

\def\b{{\beta}}
\def\bh{{\widehat\beta}}

\def\g{{\gamma}}

\def\d{{\delta}}
\def\e{{\epsilon}}
\def\s{{\sigma}}
\def\k{{\kappa}}

\def\L{{\Lambda}}

\def\half{{1\over 2}}
\def\p{{\partial}}

\def\t{{\theta}}
\def\th{{\widehat\theta}}
\def\bar{\overline}

\Title{\vbox{\baselineskip12pt
\hbox{IFT-P.056/2004 }}}
{{\vbox{\centerline{Quantum Consistency of the Superstring }
\smallskip
\centerline{in $AdS_5\times S^5$ Background}}} }
\bigskip\centerline{Nathan Berkovits\foot{e-mail: nberkovi@ift.unesp.br}}
\bigskip
\centerline{\it Instituto de F\'\i sica Te\'orica, Universidade Estadual
Paulista}
\centerline{\it Rua Pamplona 145, 01405-900, S\~ao Paulo, SP, Brasil}

\vskip .3in
Using arguments based on BRST cohomology,
the pure spinor formalism for the superstring in an $AdS_5\times S^5$ 
background is proven to be BRST invariant and conformally invariant
at the quantum level to all orders in perturbation theory.
Cohomology arguments are also used to prove
the existence of an infinite set of non-local BRST-invariant charges
at the quantum level.

\vskip .3in

\Date {November 2004}

\newsec{Introduction}

The superstring worldsheet action
in an $AdS_5\times S^5$ Ramond-Ramond background can be studied at
the classical level using either the Green-Schwarz (GS) formalism \ref
\gs{M.B. Green and J.H. Schwarz, {\it Covariant Description of 
Superstrings}, Phys. Lett. 131B (1984) 367.} or
the pure spinor formalism \ref\pure {N. Berkovits,
{\it Super-Poincar\'e Covariant Quantization of the Superstring},
JHEP 0004 (2000) 018, hep-th/0001035.}.
Since the $AdS_5\times S^5$ Ramond-Ramond background
is a solution of Type IIB supergravity, the worldsheet action 
is classically $\k$-invariant in the GS formalism 
\ref\metsaev{R.R. Metsaev and A.A. Tseytlin
{\it Type IIB Superstring Action in $AdS_5\times S^5$ Background},
Nucl. Phys. B533 (1998) 109, hep-th/9805028.} and is
classically BRST-invariant in the pure spinor formalism
\ref\pureads{N. Berkovits and O. Chand\'{\i}a, 
{\it Superstring Vertex Operators in an $AdS_5\times S^5$ Background},
Nucl. Phys. B596 (2001) 185, hep-th/0009168.}.
In both these formalisms for the superstring, an
infinite set of non-local classically conserved
charges has been constructed which might be related to integrability
\ref\Bena{I. Bena, J. Polchinski and R. Roiban, {\it Hidden Symmetries
of the $AdS_5\times S^5$ Superstring}, Phys. Rev. D69 (2004) 046002,
hep-th/0305116.}\ref\Vallilo{
B.C. Vallilo,
{\it Flat Currents in the Classical $AdS_5\times S^5$ Pure Spinor
Superstring}, JHEP 0403 (2004) 037, hep-th/0307018.}.
At the classical level, these non-local
charges have been shown to be $\k$-invariant
in the GS formalism and BRST-invariant in the pure spinor formalism \ref\brst
{N. Berkovits, {\it BRST Cohomology and Nonlocal Conserved Charges},
hep-th/0409159}.

For applications to the AdS-CFT conjecture, it is important to know
if the worldsheet action and non-local charges
remain $\k$-invariant or BRST-invariant after including quantum corrections.
Because of quantization problems in the GS formalism, 
quantum $\k$-invariance is difficult to discuss, except 
perhaps near the plane-wave
limit in which light-cone GS methods can be used \ref\callan
{C.G. Callan Jr., H.K. Lee, T. McLoughlin, J.H. Schwarz, I. Swanson
and X. Wu, {\it Quantizing String Theory in $AdS_5\times S^5$: Beyond the
PP Wave}, Nucl. Phys. B673 (2003) 3, hep-th/0307032\semi
C.G. Callan Jr., T. McLoughlin and I. Swanson, {\it Higher
Impurity ADS/CFT Correspondence in the Near-BMN Limit},
Nucl. Phys. B700 (2004) 271, hep-th/0405153\semi
I. Swanson, {\it Quantum String Integrability and AdS/CFT}, hep-th/0410282.}.
However, since some
isometries of the $AdS_5\times S^5$
background are not manifest near the plane-wave limit,
computations using this light-cone GS method appear quite complicated.

Using the pure spinor formalism for the superstring, there are no problems
with quantization and one can
easily discuss BRST invariance at the quantum
level. In this paper, it will be proven using cohomology
arguments that the BRST transformation
of the quantum worldsheet
effective action in an $AdS_5\times S^5$ background can
be cancelled by adding a local counterterm. 
The proof relies on the algebraic renormalization method of \ref\sorella
{O. Piguet and S.P. Sorella, {\it Algebraic Renormalization:
Perturbative Renormalization, Symmetries and Anomalies}, Lect. Notes
Phys. M28 (1995) 1\semi
G. Barnich. F. Brandt and M. Henneaux, {\it Local BRST Cohomology in
Gauge Theories}, Phys. Rept. 338 (2000) 439, hep-th/0002245.} in which
trivial BRST cohomology at ghost-number $+1$ implies quantum BRST
invariance to all orders in perturbation theory. 

Furthermore, it will be proven that after adding this local counterterm,
the quantum worldsheet action is conformally invariant to all
orders in perturbation theory. This proof uses a $U(2,2|4)$-invariant
generalization of the worldsheet action and is similar to Witten's proof
of quantum 
conformal invariance in \ref\vw{N. Berkovits, C. Vafa and E. Witten, 
{\it Conformal Field Theory of AdS Background with Ramond-Ramond Flux},
JHEP 9903 (1999) 018, hep-th/9902098.}
for the superstring in an $AdS_3\times S^3$ Ramond-Ramond background.
Note that it was previously shown by explicit computation that the 
worldsheet
action in an $AdS_5\times S^5$ background is conformally invariant at one-loop
in $\a'$ \ref\valtwo{N. Berkovits, M. Bershadsky, T. Hauer, S. Zhukov and
B. Zwiebach, {\it Superstring Theory on $AdS_2\times S^2$ as a Coset
Supermanifold}, Nucl. Phys. B567 (2000) 61, hep-th/9907200\semi
B.C. Vallilo, {\it One-Loop Conformal Invariance of the Superstring
in an $AdS_5\times S^5$ Background}, JHEP 0212 (2002) 042, hep-th/0210064.}.
And it was argued based on isometries that the $AdS_5\times S^5$
background is not modified by higher-derivative corrections to the
supergravity equations of motion \ref\kallosh{R. Kallosh and A. Rajaraman,
{\it Vacua of M Theory and String Theory}, Phys. Rev. D58 (1998)
125003, hep-th/9805041}.


In a recent paper, it was proven that whenever certain
ghost-number $2$ states are absent from the BRST cohomology, one
can construct an infinite set of non-local BRST-invariant charges. 
It will be shown here that the ghost-number $2$ cohomology is
trivial in an $AdS_5\times S^5$ background, implying the existence
of an infinite set of non-local BRST-invariant charges at the quantum
level.

At the classical level, these non-local BRST-invariant charges were shown
in \brst\ to coincide with the classically conserved non-local charges
found by Vallilo \Vallilo. 
To explicitly construct the quantum non-local charges,
one would first need to compute the quantum effective action. This computation
is currently being done to one-loop order in collaboration with Brenno
Carlini Vallilo \ref\prog{N. Berkovits and B.C. Vallilo, work in progress.},
and some formulas in this paper have come from that
collaboration. Although it is not obvious that quantum BRST invariance
of the charges will automatically imply quantum
conservation, it is reasonable to assume that
BRST-invariant charges of zero ghost-number in the
pure spinor formalism necessarily commute with the Hamiltonian.\foot{In 
string theory, one usually assumes
that any BRST-invariant operator of zero ghost-number can
be put into Siegel gauge by adding an appropriate BRST-trivial operator.
Siegel gauge implies that the operator commutes with the zero mode of 
the $b$ ghost, so BRST-invariant operators in Siegel gauge
commute with the Hamiltonian
$H= \{Q,b_0\}$. In the pure spinor formalism,
there are no operators of negative ghost number, so
there are no BRST-trivial operators
of zero ghost number and there is no natural $b$ ghost. 
It therefore appears that Siegel
gauge is automatically imposed on ghost-number zero operators
in the pure spinor formalism, implying that BRST-invariant charges
of ghost-number zero necessarily
commute with the Hamiltonian. }

In section 2 of this paper, the classical worldsheet action is reviewed
using the pure spinor formalism for the superstring in
an $AdS_5\times S^5$ background. After adding appropriate local counterterms,
the quantum worldsheet effective action is proven to be
$SO(4,1)\times SO(5)$ gauge-invariant in section 3,
BRST invariant in section 4, and conformally
invariant in section 5. Finally, in section 6, an infinite set of
non-local BRST-invariant currents are proven to exist at the quantum level.

\newsec{Review of Pure Spinor Formalism in $AdS_5\times S^5$ Background}

In this section, the classical worldsheet
action in an $AdS_5\times S^5$ background is reviewed
using the pure spinor formalism
for the superstring. 
As in the Metsaev-Tseytlin 
GS action in an $AdS_5\times S^5$ background
\ref\metsaev{R.R. Metsaev and A.A. Tseytlin
{\it Type IIB Superstring Action in $AdS_5\times S^5$ Background},
Nucl. Phys. B533 (1998) 109, hep-th/9805028.},
the action
in the pure spinor formalism \pureads\ is constructed from left-invariant
currents $J^A = (g^{-1} \p g)^A$ where
$g(x,\t,\widehat\t)$ 
takes values in the coset $PSU(2,2|4)/(SO(4,1)\times SO(5))$,
$A= ([ab],m,\a,\ah)$ ranges over the
30 bosonic and 32 fermionic elements in the Lie algebra of $PSU(2,2|4)$,
$[ab]$ labels the $SO(4,1)\times SO(5)$ ``Lorentz''
generators, $m=0$ to 9 labels
the ``translation'' generators, and $\a,\ah=1$ to 16 label the 
fermionic ``supersymmetry'' generators.
The action in the pure spinor formalism also involves left and
right-moving bosonic ghosts, $(\l^\a, w_\a)$ and $(\lh^\ah,\wh_\ah)$,
which satisfy the pure spinor constraints
$\l\g^m\l=\lh\g^m\lh=0$. Because of the pure spinor constraints,
$w_\a$ and $\widehat w_\ah$ can only appear in combinations which are
invariant under $\d w_\a= \xi^m (\g_m\l)_\a$ and
$\d \widehat w_\ah= \widehat\xi^m (\g_m\lh)_\ah$.
These pure spinor ghosts couple to the
$AdS_5\times S^5$ spin connection $J^{[ab]}$
in the worldsheet action through
their Lorentz currents
$N_{ab} = \half w\g_{ab}\l$ and $\widehat N_{ab} = \half \wh\g_{ab}\lh$.

Using the notation defined below, the
classical worldsheet action is
\eqn\class{S_0 = \langle \half J_2 \bar J_2 + {3\over 4}
J_3 \bar J_1  + {1\over 4} 
J_1 \bar J_3
+ 
w  \bar\nabla \lambda  + 
\widehat w \nabla \lh  - N \widehat N \rangle}
\eqn\classtwo{= 
\langle \half (J_2 \bar J_2 + J_3 \bar J_1 + J_1 \bar J_3) +
{1\over 4}(J_3\bar J_1 - J_1\bar J_3) 
+ ( w\bar\p\l + \widehat w\p\lh + N\bar J_0 + \widehat N J_0 -
N \widehat N)\rangle }
where
\eqn\notat{J_0 = (g^{-1}\p g)^{[ab]} T_{[ab]},\quad
J_1 = (g^{-1}\p g)^{\a} T_{\a},\quad
J_2 = (g^{-1}\p g)^{m} T_{m},\quad
J_3 = (g^{-1}\p g)^{\ah} T_{\ah},}
$$
w = w_\a T_\ah \d^{\a\ah}, 
\quad\l = \l^\a T_\a,\quad N =  -\{w,\l\}, $$
$$\overline J_0 = (g^{-1}\overline \p g)^{[ab]} T_{[ab]},\quad
\overline J_1 = (g^{-1}\overline \p g)^{\a} T_{\a},\quad
\overline J_2 = (g^{-1}\overline \p g)^{m} T_{m},
\quad \overline J_3 = (g^{-1}\overline \p g)^{\ah} T_{\ah}, $$
$$\widehat w = \widehat w_\ah T_\a \d^{\a\ah}, 
\quad\lh = \lh^\ah T_\ah,\quad \widehat N = 
-\{\widehat w,\lh\},$$
$$\nabla Y = \partial Y + [J_0, Y], \quad
\bar\nabla Y = \bar\partial Y + [\bar J_0, Y], $$
$ \d_{\a\bh} = (\g^{01234})_{\a\bh},$
$\langle w\bar\p\l + 
\widehat w\p\lh\rangle$
is the action in a flat background for the pure spinors,
$T_A$ are the $PSU(2,2|4)$ Lie algebra generators, and
$\langle ~~\rangle$ denotes a super-trace over the $PSU(2,2|4)$ matrices
and integration over the two-dimensional worldsheet, e.g. 
$\langle J_2\bar J_2\rangle = \int d^2 z (g^{-1}\p g)^m (g^{-1}\bar\p g)^n
STr(T_m T_n).$ Note that 
\eqn\metric
{\{T_\a,T_\b\} = \g_{\a\b}^m T_m,\quad
\{T_\ah,T_\bh\} = \g_{\ah\bh}^m T_m,\quad
\{T_\a,T_\bh\} = (\half \g^{[ab]}\g^{01234})_{\a\bh} T_{[ab]},}
$${\rm and}\quad Str( T_{[ab]} T_{[cd]}) = \d_{a[c}\d_{d]b},\quad
Str( T_m T_n) = \eta_{mn},\quad
Str( T_\a T_\bh) = 
- Str( T_\bh T_\a) =  \d_{\a\bh}.$$
The action of \class\ is manifestly invariant under global $PSU(2,2|4)$
transformations which transform $g(x,\t,\th)$ by left multiplication
as $\d g = (\Sigma^A T_A) g$ and is also manifestly invariant
under local $SO(4,1)\times SO(5)$ gauge transformations which transform
$g(x,\t,\th)$ by right multiplication as $\d_\Lambda 
g= g\Lambda$ and transform
the pure spinors as 
$$\d_\L \l = [\l,\L],\quad \d_\L\lh = [\lh,\L],\quad
\d_\L w = [w,\L],\quad \d_\L\widehat w = [\widehat w,\L]$$
where $\L = \L^{[ab]} T_{[ab]}$.

Under classical
BRST transformations generated by 
$$\e Q = \e\int d\s  STr(\l J_3 + 
\lh \bar J_1)  $$
where $\e$ is a constant anticommuting parameter,
$g(x,\t,\th)$ transforms by right-multiplication as
\eqn\brstt{\e Q (g) = g (\e\l+\e\lh)}
and the pure spinors transform as
$$\e Q(w) = - J_3 \e,\quad
\e Q(\widehat w) = -\bar J_1 \e, \quad
\e Q(\l)= \e Q(\lh) =0,$$
which implies that
$$\e Q (N) =  [J_{3},\e\l],\quad \e Q( \widehat N )= 
 [\bar J_{1},\e\lh].$$
The left-invariant currents of \notat\ transform under \brstt\ as
$$\e Q (J_j) = \d_{j+3,0} ~\p(\e\l) + [J_{j+3},\e\l] +\d_{j+1,0}~\p(\e\lh) + 
[J_{j+1},\e\lh],$$
$$\e Q (\bar J_j )=\d_{j+3,0}~\bar\p(\e\l) + [\bar J_{j+3},\e\l] +
\d_{j+1,0}~\bar\p(\e\lh) +
 [\bar J_{j+1},\e\lh],$$
where $j$ is defined modulo 4, i.e. $J_j \equiv J_{j+4}$.

One can easily verify that $S_0$ is the unique $PSU(2,2|4)$-invariant
expression which is BRST invariant under \brstt. To verify this,
note that the first term in
\classtwo\ transforms under \brstt\ to
$$\half\langle J_3 \bar\nabla(\e\lambda) +\bar J_3 \nabla (\e\lambda)
+ J_1 \bar\nabla(\e\lh) + \bar J_1 \nabla(\e\lh)\rangle.$$ 
Using the
Maurer-Cartan equations
\eqn\mceq{\nabla \bar J_3 - \bar\nabla J_3 = -[J_1,\bar J_2] -[J_2,\bar J_1],
\quad
\nabla \bar J_1 - \bar\nabla J_1 = -[J_3,\bar J_2] -[J_2,\bar J_3],}
the second term in \classtwo\ transforms under \brstt\ to
$$\half\langle J_3 \bar\nabla(\e\lambda) -\bar J_3 \nabla(\e \lambda)
- J_1\bar\nabla(\e\lh) + \bar J_1 \nabla(\e\lh)\rangle.$$ And the last term
in \classtwo\ transforms under \brstt\ to
$$\langle - J_3 \bar\nabla(\e\lambda) -
\bar J_1 \nabla(\e\lh)\rangle.$$

\newsec{$SO(4,1)\times SO(5)$ Gauge Invariance}

Using the classical BRST transformation of \brstt\ 
and $\{\l,\l\} =\{\lh,\lh\}=0$
from the pure spinor constraints, one finds that
\eqn\nilp{Q^2 (g) = - g \{\l,\lh\}, }
$$Q^2 (N) = 
 - [N, \{\l,\lh\}] -  \{\l, \nabla \lh - [N,\lh]\},$$
$$
Q^2 (\widehat N) = 
 - [\widehat N, \{\l,\lh\}] - \{\lh, \bar\nabla \l - [\widehat N,\l]\}.$$
Furthermore, $[\l,\{\l,\lh\}]=
[\lh,\{\l,\lh\}]= 0$ implies that 
\eqn\nilpl{Q^2 (\l) = 0 = - [\l,\{\l,\lh\}],\quad
Q^2 (\lh) = 0 =  -[\lh,\{\l,\lh\}].}

So up to an $SO(4,1)\times SO(5)$
gauge transformation parameterized by 
\eqn\paramg{\{\l,\lh\} = \l^\a \lh^\bh (\half
\g^{[ab]}\g^{01234})_{\a\bh} T_{[ab]} ,}
and up to the classical equations of motion
\eqn\eql{ \nabla \lh - [N,\lh]=0\quad \quad{\rm and}\quad\quad
\bar\nabla \l - [\widehat N,\l] =0,}
$Q$ is nilpotent. Since the classical action of \class\ is invariant
under 
$SO(4,1)\times SO(5)$
gauge transformations,
$Q$ is therefore a consistent BRST transformation
at the classical level.

It will now be argued that after adding a local counterterm, the quantum
effective action remains invariant under 
$SO(4,1)\times SO(5)$
gauge transformations. This is essential for consistency of the
BRST transformation at the quantum level.
To prove that such a local counterterm can always be found, note that 
the $SO(4,1)\times SO(5)$ gauge transformation of the quantum effective
action, $\d_\Lambda S_q$, must be a local operator since any quantum anomaly
comes from a short-distance regulator. Furthermore, since global
$SO(4,1)\times SO(5)$ invariance is manifest, $\d_\Lambda S_q$ must
vanish
when the $SO(4,1)\times SO(5)$ gauge parameter $\Lambda = \Lambda^{[ab]}
T_{[ab]}$ is constant. Therefore, 
\eqn\sofour{\d_\Lambda S_q = \int d^2 z (f_{[ab]}\bar\partial \Lambda^{[ab]}
+\bar f_{[ab]}\partial \Lambda^{[ab]})}
where $f_{[ab]}$ and $\bar f_{[ab]}$ are some 
operators which carry $(left,right)$
conformal weight $(1,0)$ and $(0,1)$ respectively. 

The only candidates for $f_{[ab]}$ and $\bar f_{[ab]}$ are 
$(N_{[ab]},J_{[ab]})$ and
$(\bar N_{[ab]},\bar J_{[ab]})$, so
\eqn\firsts{\d_\Lambda S_q = 
\langle c_1 N\bar\partial \Lambda
+\bar c_1  \widehat N\partial \Lambda
+ c_2  J_0\bar\partial \Lambda
+\bar c_2 \bar J_0\partial \Lambda\rangle}
for some constants $(c_1,\bar c_1,c_2,\bar c_2)$. 
By adding the local counterterm
\eqn\counterf{S_c = -\langle c_1 N\bar J_0 +
  \bar c_1 \widehat N J_0
+\half (c_2+\bar c_2) 
  J_0 \bar J_0\rangle,}
one can cancel most of the variation to obtain
\eqn\counters{
\d_\Lambda (S_q + S_c) = 
\half( c_2-\bar c_2) \langle J_0\bar\partial \Lambda
- \bar J_0\partial \Lambda\rangle),}
which is the standard parity-violating anomalous variation in two dimensions.

Although the worldsheet action of \class\ is not invariant under a parity
transformation which exchanges $z$ with $\bar z$,
the action is invariant under a transformation which simultaneously
exchanges
$z$ with $\bar z$, $\l$ with $\lh$, $w$ with $\widehat w$, and $\t$ with
$\th$. This implies that $c_1=\bar c_1$ and $c_2=\bar c_2$ in
$\d_\Lambda S_q$ in \firsts. So the anomalous variation of \counters\
vanishes, implying that $\d_\Lambda(S_q + S_c)=0$.

\newsec{Quantum BRST Invariance}

In this section, trivial BRST cohomology at ghost-number $+1$ will be
used to prove that the BRST transformation of the quantum effective
action $S_q$ can be cancelled by adding a local counterterm. 
Since the BRST
transformation of \brstt\ commutes with $SO(4,1)\times SO(5)$ gauge
transformations and since $\d_\Lambda S_q =0$ after adding the
counterterm of the previous section, the BRST transformation of $S_q$
satisifes $\d_\Lambda Q (S_q)=0$.
Furthermore, the BRST variation of the quantum effective action
must be a local operator since quantum anomalies come from a short-distance
regulator. 

So $Q(S_q)$ is an $SO(4,1)\times SO(5)$ gauge-invariant
local operator of ghost-number $+1$, which implies it can be written as
\eqn\brsts{\e Q(S_q) = \langle a_1 \bar J_2 [J_3, \e\lh]
+ \bar a_1
J_2 [\bar J_1, \e\l]
+a_2 \bar J_2 [J_1, \e\l]
+ \bar a_2
 J_2 [\bar J_3, \e\lh]}
$$
+a_3 J_3 [\widehat N, \e\l]
+ \bar a_3
 \bar J_1 [N, \e\lh] 
+ a_4 
 J_3 \bar \nabla (\e\l) 
+ \bar a_4 
\bar J_1 \nabla(\e \lh) \rangle $$
where $a_j$ and $\bar a_j$
are some constants.\foot{
Terms such as $\langle \bar J_2 \{J_3, \e\lh\}\rangle$ do not
need to be considered since the effective action (e.g., using
the background field method) and BRST transformations
only involve the structure constants $f_{AB}^C$
and do not involve constants such as $d_{AB}^C$ coming from 
anticommutators.}
Note that $[\widehat N,\e\lh]=
[N,\e\l]=0$ because of the pure spinor constraint and that terms such as
$\langle \bar J_3\nabla(\e\l)\rangle$ can be related to terms
in \brsts\ by integrating by parts and using the Maurer-Cartan equation
$\nabla \bar J_3 - \bar\nabla J_3 = [\bar J_1, J_2] + [\bar J_2, J_1].$

Since $Q$ is nilpotent on $SO(4,1)\times SO(5)$
gauge-invariant operators up to the equations of \eql, 
$Q^2(S_q)$ must be proportional to the equations of \eql.
Using the BRST transformations of \brstt, this implies that the
coefficients in \brsts\ must satisfy\foot{I would like to thank
Brenno Vallilo for discussions on this computation.}
\eqn\mustsat{a_1 =\bar a_1,\quad a_2 = \bar a_2,\quad a_3 +a_4=\bar a_3+
\bar a_4.}

It will now be shown that whenever the restriction of \mustsat\ is satisfied, 
$Q(S_q)$ can be written as the BRST variation of a local counterterm. In 
other words, the BRST cohomology of local ghost-number $+1$ operators
is trivial. Using the BRST transformations of \brstt, one finds
that the local counterterm $S_c$ which satisfies 
$Q(S_c)= - Q(S_q)$ is
\eqn\localc{S_c = \langle - a_2 \bar J_2 J_2 +
(a_1 -a_2) \bar J_1 J_3
+ (a_3 -\bar a_4+a_2-a_1)  N\widehat N}
$$
+(a_4 + a_1 - a_2) w\bar\nabla \l
+(\bar a_4 + a_1 - a_2)  \widehat w\nabla \lh\rangle .$$
So $Q(S_q + S_c)=0$, implying that the quantum effective action
$S_{eff}=S_q + S_c$ is
invariant under BRST transformations.

Using the algebraic renormalization method of \sorella, this proof of
quantum BRST invariance can be extended by induction
to all orders in perturbation
theory.\foot{Although the algebraic renormalization method of \sorella\
uses the ``gauge-invariant'' BRST cohomology including antifields, the proof
here uses the ``gauge-fixed'' BRST cohomology where antifields have been
set to zero. As discussed in \ref\henn{
M. Henneaux, {\it
On the Gauge-Fixed BRST Cohomology}, Phys. Lett. B367 (1996) 163,
hep-th/9510116\semi
G. Barnich, M. Henneaux, T. Hurth and 
K. Skenderis, {\it Cohomological Analysis of Gauge-Fixed Gauge Theories},
Phys. Lett. B492 (2000) 376, hep-th/9910201\semi
G. Barnich, T. Hurth and 
K. Skenderis, {\it Comments on the Gauge-Fixed BRST Cohomology
and the Quantum Noether Method}, Phys. Lett. B588 (2004) 111, 
hep-th/0306127.}, the gauge-fixed cohomology
is sufficient for proving quantum BRST invariance if quantum modifications
to the gauge-fixed BRST operator can be defined such that nilpotence
is preserved. This is possible if there are no conserved currents of
ghost-number $2$ which could deform $Q^2$. A counter-example discussed in
\henn\
is the conserved current $j_\mu = C \p_\mu C$ in Maxwell theory where
$C$ is the fermionic ghost whose equation of motion in Lorentz gauge is
$\p_\mu\p^\mu C=0$. Fortunately, one can easily check that there are no
conserved currents of ghost-number $2$ for the action of 
\class, so the gauge-fixed BRST cohomology is sufficient for proving
quantum BRST invariance.}
For example, suppose the quantum effective action is BRST invariant
up to order $h^{n-1}$, i.e. $\widetilde
Q(S_{eff}) = h^n \Omega +{\cal O}(h^{n+1})$ for some local $\Omega$ where
$\widetilde Q = Q + Q_q$ and $Q_q$ generates quantum corrections to the
classical BRST transformations of \brstt\ generated by $Q$. Since $Q\Omega=0$,
trivial cohomology at ghost-number $+1$ implies that
$\widetilde Q(S_{eff} - h^n \Sigma)= {\cal O}(h^{n+1})$ where 
$\Sigma$ is a local operator satisfying $Q\Sigma=\Omega$. 
So the quantum effective action $S_{eff}-h^n\Sigma$ is
BRST invariant up to order $h^n$.

\newsec{Quantum Conformal Invariance}

To prove that the quantum effective action is conformally invariant, 
a trick shall be used which was previously used for the superstring
in an $AdS_3\times S^3$ Ramond-Ramond background \vw. The trick is to enlarge
the $PSU(2,2|4)$ Lie algebra to a $U(2,2|4)$ Lie
algebra. In other words, include two new bosonic generators, $I$ and $L$,
satisfying the commutation relations
\eqn\newcomm{[L, T_\a] = \d_\a^\bh T_\bh,\quad
[L, T_\ah] = - \d_\ah^\b T_\b,}
$$\{T_\a, T_\b\} = \g^m_{\a\b} T_m + (\g^{01234})_{\a\b} I,\quad
\{T_\ah, T_\bh\} = \g^m_{\ah\bh} T_m + (\g^{01234})_{\ah\bh} I.$$
So the $U(2,2|4)$ generators $(I,L,T_A)$ satisfy the algebra
\eqn\algnew{[L,T_A]= c_A^B T_B, \quad [T_A,T_B\}= f_{AB}^C T_C + d_{AB} I,
\quad [I,T_A] = [I,L]=0,}
where $f_{AB}^C$ are the $PSU(2,2|4)$ structure constants, 
$c_\a^\bh=\d_\a^\bh$, $c_\ah^\b= -\d_\ah^\b$, 
$d_{\a\b}= \g^{01234}_{\a\b} $
and 
$d_{\ah\bh}= \g^{01234}_{\ah\bh} $.
Note that $L$ acts as an outer automorphism of $PSU(2,2|4)$ 
and $I$ acts as a central extension.

Now define left-invariant currents
\eqn\kdefin{K= h^{-1} \p h, \quad
\bar K= h^{-1}\bar\p h, }
where 
$h(x,\t,\th,u,v)$ takes values in the coset
$U(2,2|4)/(SO(4,1)\times SO(5))$ and $(u,v)$ are two additional
bosonic variables which are not present in the coset
$PSU(2,2|4)/(SO(4,1)\times SO(5))$.
It is convenient to parameterize
\eqn\param{h(x,\t,\th,u,v) = \exp(u I + v L)~ g(x,\t,\th)}
where $g(x,\t,\th)$ takes values in
$PSU(2,2|4)/(SO(4,1)\times SO(5))$.
So 
\eqn\morekdef{h^{-1}\p h = K_I  + K_L  + K_0 + K_1 + K_2 + K_3 \quad 
{\rm where}}
$$K_I = (h^{-1}\p h)^I I = (\p u + (g^{-1}(\p v ~L) g)^I +(g^{-1}\p g)^I) I,$$
$$K_L = (h^{-1}\p h)^L L = (\p v) L , $$
$$K_0 + K_1 + K_2 + K_3 = 
(h^{-1}\p h)^A T_A = 
((g^{-1}(\p v ~L) g)^A  + 
(g^{-1}\p g)^A ) T_A  .$$

Under the BRST transformation 
$\e Q'(h) = h(\e \l +\e\lh)$, the left-invariant currents transform as
\eqn\newbrst{\e Q'(K_I) = [K_3,\e\lh] + [K_1,\e \l], \quad \e Q'(K_L)=0,}
$$\e Q' (K_j) = \d_{j+3,0} (\e\p\l + [K_L,\e\lh]) + [K_{j+3},\e\l] +\d_{j+1,0}
(\e\p\lh + [K_L,\e\l]) + [K_{j+1},\e\lh],$$
where $j$ is defined modulo 4, i.e. $K_j \equiv K_{j+4}$.

Now consider the classical worldsheet action
\eqn\uaction{S' = \half \int d^2 z Str(K_I\bar K_L + \bar K_I K_L) + S'_0}
where
$S'_0$ is the classical action of \class\ with $J_A$ replaced by $K_A$.
Note that $S'$ is manifestly invariant under global $U(2,2|4)$
transformations which transform $h$ by left multiplication, 
and differs from $S_0$ because of its dependence
on the two additional bosons $u$ and $v$.
Using the $U(2,2|4)$ Maurer-Cartan equations, 
\eqn\mcueq{\nabla \bar K_3 - \bar\nabla K_3 = -[K_1,\bar K_2] -[K_2,\bar K_1]
-[K_L,\bar K_1] -[K_1,\bar K_L],}
$$
\nabla \bar K_1 - \bar\nabla K_1 = -[K_3,\bar K_2] -[K_2,\bar K_3]
-[K_L,\bar K_3] -[K_3,\bar K_L] ,$$
one can check that $S'$
is invariant under the BRST transformations of \newbrst.
Furthermore, one can repeat the arguments of sections 3 and 4 to show
that the BRST transformation of the quantum effective action $S'_q$
can be cancelled by adding a local counterterm $S'_{c}$ to obtain
a BRST-invariant quantum action $S'_{eff} = S'_q + S'_{c}$.

{}From the definitions in \morekdef\ for the left-invariant currents, 
$$S' = S_0 + \int d^2 z (\p u \bar \p v + 
j(x,\t,\th)\bar\p v +  
\bar j(x,\t,\th)\p v +  k(x,\t,\th) \p v \bar \p v)$$ 
where $S_0$, $j$, $\bar j$ and $k$ are independent of $u$ and $v$.
Since there are no terms in $S'$ which are quadratic in $u$, there
is no $\langle v v\rangle$ propagator in the Feynman rules
for the quantum 
effective action. So $v$-independent terms in the quantum effective
action $S'_{eff}$ are the same as in the original $PSU(2,2|4)$-invariant
quantum effective action $S_{eff}$ of section 4.  
It will now be proven that $S'_{eff}$ is conformally invariant, which
immediately implies that $S_{eff}$ is conformally invariant since
$S'_{eff}|_{v=0} = S_{eff}$.

To prove that $S'_{eff}$ is conformally invariant, first note that
cohomology arguments imply that
quantum conformal transformations can be defined to commute with
quantum BRST transformations.\foot{To prove this, suppose that
$[\widetilde Q, \d_C] = h^n \d' + {\cal O}(h^{n+1})$
where $\widetilde Q = Q+Q_q$, $Q_q$ generates quantum corrections to
the classical BRST transformations generated by $Q$, 
$\d_C$ is the conformal transformation to 
order $h^{n-1}$, and $\d'$ is some local transformation carrying
$+1$ ghost number. Then $\{Q, \d'\}=0$ implies that 
$\d' = -[Q, \d_q]$ for some $\d_q$ because of trivial BRST cohomology
for local charges of $+1$ ghost number. So $\d_C + h^n \d_q$
can be defined as the conformal transformation to order $h^n$, and satisfies
$[\widetilde Q, \d_C + h^n \d_q]= {\cal O}(h^{n+1}).$}
So if $\d_C$ denotes the quantum
conformal transformation, $Q' (\d_C S'_{eff}) =0$. So the
conformal transformation of $S'_{eff}$ must be BRST invariant, and
must be local since it comes from a short-distance regulator.
But one can easily verify that the only $U(2,2|4)$-invariant BRST-invariant
local operator of ghost-number zero is the classical action $S'$ of \uaction.
So $\d_C S'_{eff}$ must be proportional to $S'$.
But the term $\int d^2 z \p u \bar\p v$ in $S'$ cannot receive
quantum corrections since $S'$ contains no terms quadratic in $u$.
So the term $\int d^2 z \p u \bar\p v$ cannot appear in $\d_C S'_{eff}$,
which implies that $\delta_C S'_{eff}=0$.
Note that
as in the proof of quantum
BRST invariance in section 4, this proof of quantum conformal invariance
is valid to all orders in perturbation theory. 

\newsec{Non-Local BRST-Invariant Charges}

As discussed in \brst, the existence of non-local BRST-invariant
charges in string theory is related to the triviality of a certain
BRST cohomology class. To understand this relation, consider the non-local
integrated operator 
\eqn\kdef{k^{C} = f^C_{AB} \int_{-\infty}^\infty d\s ~j^A(\s)~ 
\int_{-\infty}^\s d\s' ~j^B(\s') }
where $\int_{-\infty}^\infty d\s j^A(\s)$ are the
Noether charges for the global symmetry algebra and $f_{AB}^C$ are the
structure constants. Since the Noether charges are BRST-invariant,
$Q(j^A(\s))= \p_\s h^A(\s)$ for some $h^A(\s)$ of $+1$ ghost-number,
which implies that $Q(k^C) = -2 f_{AB}^C \int_{-\infty}^{\infty}
d\s h^A(\s) j^B(\s)$ is
a local integrated operator of $+1$ ghost-number.

Whenever $Q(k^C)$ can be written as the BRST variation of a local
integrated operator, i.e. whenever $Q(k^C) = Q(\int_{-\infty}^\infty
d\s \Sigma^C(\s))$ for some local $\Sigma^C(\s)$, one can construct
the non-local BRST-invariant charge
\eqn\qdef{q^{C} = f^C_{AB} \int_{-\infty}^\infty d\s ~j^A(\s)~ 
\int_{-\infty}^\s d\s' ~j^B(\s') ~- \int_{-\infty}^\infty d\s \Sigma^C(\s) .}
Furthermore, by repeatedly commuting $q^C$ with $q^D$, one generates
an infinite set of non-local BRST-invariant charges.

So if the BRST cohomology is trivial for local integrated operators
of $+1$ ghost-number transforming in the adjoint representation, 
one can construct an infinite set of non-local BRST-invariant charges.
Furthermore, one can use arguments similar to those of section 4 to prove
that this construction is valid at the quantum level to all orders
in perturbation theory. For example, suppose that a non-local 
BRST-invariant charge $q^A$ has been constructed to order $h^{n-1}$,
i.e. $\widetilde Q(q^C) = h^n \Omega^C +{\cal O}(h^{n+1})$ 
where $\Omega^C$ is some 
integrated operator of ghost-number $+1$, $\widetilde Q = Q + Q_q$,
and $Q_q$ generates quantum corrections to the classical BRST
transformations of \brstt\ generated by $Q$. Like
other types of quantum anomalies,
$\Omega^C$ must be a local integrated
operator since it comes from a short-distance
regulator in the operator product expansion $j^A(\s) j^B(\s')$
\ref\anom{M. Luscher, {\it Quantum Non-Local Charges and Absence of
Particle Production in the Two-Dimensional Non-Linear $\s$-Model},
Nucl. Phys. B135 (1978) 1\semi
E. Abdalla, M. Gomes and M. Forger, {\it On the Origin of Anomalies
in the Quantum Non-Local Charge for the Generalized Non-Linear
Sigma 
Models}, Nucl. Phys. B210 (1982) 181.}. 
So trivial cohomology implies that there exists a local
operator $\Sigma^C(\s)$ such that $\Omega^C= Q(\int_{-\infty}^\infty
d\s\Sigma^C(\s))$. Therefore, $q^C - h^n \int_{-\infty}^\infty
d\s\Sigma^C(\s)$ is BRST-invariant to order $h^n$.

To verify that the relevant cohomology class is trivial for the superstring
in an $AdS_5\times S^5$ background, it will be useful to recall that
for every integrated operator of ghost-number $+1$ in the
BRST cohomology, there exists a corresponding unintegrated operator
of ghost-number $+2$ and zero conformal weight in the BRST cohomology.
This is easy to prove since $Q(\int d\s W(\s))=0$ implies that
$Q(W(\s)) = \p_\s V(\s)$ where $V(\s)$ is a BRST-invariant operator
of zero conformal weight. And if $V$ is BRST-trivial, i.e. if
$V=Q\Lambda$ for some $\Lambda$, then $Q(W-\p_\s\Lambda)=0.$
Since the BRST cohomology is trivial for unintegrated operators of
nonzero conformal weight, $W-\p_\s\Lambda= Q\Sigma$ for some $\Sigma$.
So 
\eqn\fineq{\int d\s W(\s) = \int d\s (Q\Sigma(\s) + \p_\s\Lambda(\s)) =
Q(\int d\s \Sigma(\s)),}
which implies that $\int d\s W(\s)$ is BRST-trivial \ref\bigp{N. Berkovits,
M.T. Hatsuda and W. Siegel, {\it The Big Picture}, Nucl. Phys. B371 (1992)
434, hep-th/9108021.}.

At ghost-number $+2$, the only unintegrated operators of zero conformal
weight which transform in the adjoint representation of the global
$PSU(2,2|4)$ algebra are
\eqn\twoun{V_1 = g~(\l^\a T_\a)(\lh^\bh T_\bh)~ g^{-1} \quad
{\rm and} \quad
V_2 = g~(\lh^\ah T_\ah)(\l^\b T_\b)~ g^{-1} ,}
where $g(x,\t,\th)$ transforms by left multiplication as
$\d g(x,\t,\th)= \Sigma g(x,\t,\th)$ under the global
$PSU(2,2|4)$ transformation parameterized by $\Sigma=\Sigma^A T_A$.
One can easily verify that $Q(V_1-V_2)\neq 0$ and that
$V_1 + V_2 = Q\Omega$ where 
\eqn\veriom{
\Omega = \half g~ (\l^\a T_\a + \lh^\ah T_\ah)~ g^{-1}.}
So the cohomology is trivial, which implies the existence of an infinite
set of non-local BRST-invariant charges at the quantum level.

\vskip 15pt
{\bf Acknowledgements:} I would like to thank  
Yoichi Kazama, Yutaka Matsuo, Silvio Sorella, Brenno Carlini Vallilo
and Edward Witten for useful discussions,
CNPq grant 300256/94-9, 
Pronex 66.2002/1998-9,
and FAPESP grant 99/12763-0
for partial financial support, and the Funda\c{c}\~ao Instituto de
F\'{\i}sica Te\'orica 
for their hospitality.

\listrefs

\end